\documentclass[submission,copyright,creativecommons]{eptcs}
\providecommand{\event}{Non-Classical Logic. Theory and Applications} % Name of the event you are submitting to
\usepackage{breakurl}             % Not needed if you use pdflatex only.
\usepackage{underscore}           % Only needed if you use pdflatex.
\usepackage{amssymb}
\usepackage{multicol}

\title{Dunn Semantics for Contra-Classical Logics}
\author{Luis Estrada-Gonz\'alez
\institute{Institute for Philosophical Research\\
National Autonomous University of Mexico\thanks{This paper was written during the COVID-19 crisis. A previous version was presented at the I CILEM (Costa Rica). I want to thank Axel Arturo Barceló-Aspeitia, Lorenzo Boccafogli, Fernando Cano-Jorge, Ricardo Arturo Nicolás-Francisco, Hitoshi Omori, Elisángela Ramírez-Cámara, Christian Romero and many, many others for useful discussions closely related to the present topic. I want to thank the support from the following sources: DGAPA-UNAM through a PASPA sabbatical grant and the PAPIIT project IG400422; the Notre Dame International-Mexico Faculty Grant Program project ``The scope and limits of non-detachable conditionals''; the Coimbra Group and the KU Leuven through a scholarship from the Programme for Young Professors and Researchers from Latin American Universities. Finally, I thank the reviewers for their precious comments.}\\
Mexico City, Mexico}
\institute{Faculty of Physico-Mathematical Sciences\\Meritorious Autonomous University of Puebla\\
Puebla, Mexico}
\email{loisayaxsegrob@comunidad.unam.mx}
}
\def\titlerunning{Dunn Semantics for Contra-Classical Logics}
\def\authorrunning{L. Estrada-Gonz\'alez}
\begin{document}
\maketitle

\begin{abstract}
In this paper I show, with a rich and systematized diet of examples, that many contra-classical logics can be presented as variants of \textbf{FDE}, obtained by modifying at least one of the truth or falsity conditions of some connective. Then I argue that using Dunn semantics provides a clear understanding of the source of contra-classicality, namely, connectives that have either the classical truth or the classical falsity condition of another connective. This requires a fine-grained analysis of the sorts of modifications that can be made to an evaluation condition, analysis which I offer here as well.
\end{abstract}

\section{Introduction}
Said briefly, Dunn semantics is a modelling of formal languages where truth and falsity are the only truth values but they are not always related functionally to formulas. In Dunn semantics, there are  four admissible interpretations for a formula: it can be just true, just false, neither true nor false and both true and false. Dunn semantics is especially associated to the logic \textbf{FDE}, which is important because its generality has proved to be fruitful as a basis for developing further logics. There are three well-known logics that can be obtained as extensions of \textbf{FDE}, that is, by eliminating one of the admissible interpretations in the most general Dunn semantics. One of those logics is (strong) Kleene logic, \textbf{K$_{3}$}, which leaves the both true and false interpretation out; another is González-Asenjo/Priest's \textbf{LP}, which leaves the neither true nor false interpretation out. The third one is classical logic, which leaves out the two that are left out in \textbf{K$_{3}$} and \textbf{LP}.

But those are not the only logics that can be studied as variations on \textbf{FDE}. David Nelson \cite{Nelson1949} clearly distinguished between truth and falsity constructive conditions for the connectives, very much in the spirit of Dunn semantics for \textbf{FDE} to be made explicit below. It was precisely working with Nelson's logic \textbf{N4}\footnote{See \cite{KamideWansing2012} for an overview of this logic.} that Wansing \cite{Wansing2005} obtained his connexive logic \textbf{C}. He changed the falsity condition for $A\rightarrow B$ in \textbf{N4} and the resulting logic was no longer subclassical, but \emph{contra-classical}. This means that, without enriching the language, it validates arguments that are not valid in classical logic.\footnote{And, because of the Post-completeness of classical logic, this means that the resulting logic must be either trivial ---i.e. validates every argument---, which is not, or come with the invalidity of some classically valid arguments. In fact, the resulting logic is \emph{hyper-connexive}, because besides validating the core connexive schemas, namely

\noindent
$\sim \! (A\rightarrow\sim \! A)$\ \ \ \ \ \ \ \ \ \ \ \ \ \ \ \ \ \ \ \ \ \ \ \ \ Aristotle's Thesis

\noindent
$\sim \! (\sim \! A\rightarrow A)$\ \ \ \ \ \ \ \ \ \ \ \ \ \ \ \ \ \ \ \ \ \ \ \ Variant of Aristotle's Thesis

\noindent
$(A \rightarrow B) \rightarrow \sim \! (A \rightarrow \sim \! B)$\ \ \ \ \ \ Boethius' Thesis

\noindent
$(A \rightarrow \sim \! B) \rightarrow \sim \! (A \rightarrow B)$\ \ \ \ \ \ Variant of Boethius' Thesis

\noindent
it also validates the converses of Boethius' Theses.}

After Wansing, Omori \cite{Omori2016} used the same idea, modifying the falsity condition for the conditional on top of \textbf{LFI1} to get another connexive logic, \textbf{dLP}. After that, he has shown (cf. \cite{Omori20XX}) that a number of well-known and new paraconsistent and relevant logics can be obtained also by modifying appropriately the falsity condition for some connectives while leaving the \textbf{FDE}-like truth and falsity conditions for the remaining ones, in most cases even negation, fixed. More recently, Wansing and Unterhuber \cite{WansingUnterhuber2019} modified the falsity condition of Chellas' basic conditional logic \textbf{CK} and they obtained a (weakly) connexive logic. Such a general approach to logics ---viz. starting with Dunn-like evaluations for \textbf{FDE} and then modify them to obtain different logics--- has been called the `Bochum Plan'.

But Dunn semantics is not only a tool for crafting new logics, but also to provide new understandings of already existing ones. For example, Omori and Wansing \cite{OmoriWansing2019} have put forward a systematization of connexive logics based on certain controlled modifications in the conditional's truth and falsity conditions, showing that, in general, modifying the truth condition has led only to \emph{weak connexivity} (Boethius’ Theses hold only in rule form, if at all), whereas modifying the falsity condition has led to \emph{hyper-connexivity} (i.e.  not only do Boethius’ Theses hold, but also their converses). More recently, Estrada-González and Cano-Jorge \cite{EstradaCano2021} showed how a Dunn semantics can help to address some of the objections raised against Reichenbach's three-valued logic.

With this background, the aim of this paper is twofold. First, to show that many contra-classical logics can be presented as variants of \textbf{FDE}, obtained by modifying at least one of the truth or falsity conditions of some connective. Whereas this way of presenting contra-classical logics is not new, my contribution here is that the diet of examples will be enriched and systematized: examples will be given for all the possible modifications in the given language; when possible, taken from existing literature. Second, to give a precise explanation of how and why the contra-classicality is obtained. The presentation using Dunn semantics provides a clear understanding of the source of contra-classicality, namely, connectives that have either the classical truth or the classical falsity condition of another connective. This requires a fine-grained analysis of the sorts of modifications that can be made to an evaluation condition, and I provide such analysis here.

The structure of the remaining of the paper is as follows. In Section 2, I revisit, at tutorial speed, the basics of Dunn semantics for \textbf{FDE}. In Section 3, I show how to treat systematically some contra-classical logics through modifications of the truth and falsity conditions for the connectives in \textbf{FDE}. This procedure opens at least two problems. One is to explain where the contra-classicality comes from; the other is to explain whether the modified connectives are still the intended connectives and why. In Section 4, I will argue that the systematization using Dunn semantics provides a clear understanding of the source of contra-classicality, namely, connectives that have either the classical truth or the classical falsity condition of another connective. The second problem is left open and it will be tackled in a separate work.

\section{Dunn semantics and FDE}
Consider a language $L$ consisting of formulas built, in the usual way, from propositional variables with the connectives $\{\sim, \wedge, \vee, \rightarrow\}$.\footnote{The treatment of additional standard connectives, like 0-ary connectives ---$\top$, $\bot$---, modal connectives ---at least the alethic ones, $\square$, $\lozenge$--- and the usual quantifiers ---$\forall x$, $\exists x$---, is left for a future work.} I will use the first capital letters of the Latin alphabet, `$A$', `$B$', `$C$'\ldots as variables ranging over arbitrary formulas.

A key feature of Dunn semantics is that, to achieve full generality with respect to the relations between formulas and truth values, the predicates ``is true'' and ``is false'' should not be understood \emph{functionally}, that is, \emph{being true} does not imply \emph{not being false} ---nor vice versa---, and \emph{being false} does not imply \emph{not being true} ---nor vice versa---. More formally, a \emph{Dunn model} for a formal language $L$ is a relation $\sigma$ between propositional variables and values $1$ (\emph{truth}) and $0$ (\emph{falsity}), that can be extended to cover all formulas. Said otherwise, a formula can be related to the truth values, via an assignment $\sigma$, in one of the following four ways:

\begin{itemize}
\item $A$ is true but not false, represented `$1\in\sigma(A)$ and $0\notin\sigma(A)$'; more briefly, $\sigma(A)=\{1\}$
\item $A$ is true but also false, represented `$1\in\sigma(A)$ and $0\in\sigma(A)$'; more briefly, $\sigma(A)=\{1, 0\}$
\item $A$ is neither true nor false, represented `$1\notin\sigma(A)$ and $0\notin\sigma(A)$'; more briefly, $\sigma(A)=\{ \ \}$
\item $A$ is false but not true, represented `$0\in\sigma(A)$ and $1\notin\sigma(A)$'; more briefly, $\sigma(A)=\{0\}$
\end{itemize}

Now, let $\Gamma$ be a set of formulas of a logic \textbf{L}. $A$ is a \emph{logical consequence} of $\Gamma$ in \textbf{L}, $\Gamma\models_{\textbf{\tiny{L}}}A$, if and only if, for every evaluation $\sigma$, $1\in\sigma(A)$ if $1\in\sigma(B)$ for every $B\in\Gamma$. $A$ is a \emph{logical truth} in \textbf{L} if and only if $\Gamma\models_{\textbf{\tiny{L}}}A$ and $\Gamma =\varnothing$. An argument is \emph{invalid in} \textbf{L} if an only if there is an evaluation in which the premises are true, i.e. $1\in\sigma(B)$ for every $B\in\Gamma$, but the conclusion is not, i.e. $1\notin\sigma(A)$.\footnote{For simplicity, these definitions will be adapted for the other logics in this paper, just with the respective changes in language.}

\textbf{FDE} is a logic that can be presented as the result of evaluating formulas and arguments, built in the usual way from a countable set of propositional variables and the connectives $\{\sim, \wedge, \vee, \rightarrow\}$, according to the following assignments, where $A$ and $B$ stand for any formula:

\begin{itemize}
\item $\sim \! A$ is true iff $A$ is false; $\sim \! A$ is false iff $A$ is true

\item $A\wedge B$ is true iff $A$ is true and $B$ is true; $A\wedge B$ is false iff $A$ is false or $B$ is false

\item $A\vee B$ is true iff $A$ is true or $B$ is true; \ \ $A\vee B$ is false iff $A$ is false and $B$ is false

\item $A\rightarrow B$ is true iff $A$ is false or $B$ is true; $A\rightarrow B$ is false iff $A$ is true and $B$ is false
\end{itemize}

Using a Dunn model, the evaluation of formulas in  \textbf{FDE} is defined recursively as follows:

\medskip

\noindent
Either $1\in\sigma(p)$ or $1\notin\sigma(p)$, and either $0\in\sigma(p)$ or $0\notin\sigma(p)$

\noindent
$1\in\sigma(\sim \! A)$ iff $0\in\sigma(A)$

\noindent
$0\in\sigma(\sim \! A)$ iff $1\in\sigma(A)$

\noindent
$1\in\sigma(A\wedge B)$ iff $1\in\sigma(A)$ and $1\in\sigma(B)$

\noindent
$0\in\sigma(A\wedge B)$ iff either $0\in\sigma(A)$ or $0\in\sigma(B)$

\noindent
$1\in\sigma(A\vee B)$ iff either $1\in\sigma(A)$ or $1\in\sigma(B)$

\noindent
$0\in\sigma(A\vee B)$ iff $0\in\sigma(A)$ and $0\in\sigma(B)$

\noindent
$1\in\sigma(A\rightarrow B)$ iff $0\in\sigma(A)$ or $1\in\sigma(B)$

\noindent
$0\in\sigma(A\rightarrow B)$ iff $1\in\sigma(A)$ and $0\in\sigma(B)$

\medskip

\noindent
Although $A\rightarrow B$ can be defined as $\sim \! A\vee B$, considering it explicitly right from the start with a separate sign will greatly simplify the exposition. A \emph{biconditional}, $A\leftrightarrow B$, can be defined as $(A\rightarrow B)\wedge(B\rightarrow A)$.

The above model-theoretic semantics for \textbf{FDE} can be represented in a tabular way as follows:

\begin{center}
		\begin{tabular}{c|c}
			$\sim \! A$ & $A$ \\
			\hline
			$\{0\}$ 	& $\{1\}$ \\
			$\{1, 0\}$	& $\{1, 0\}$ \\
			$\{ \ \}$ & $\{ \ \}$ \\
			$\{1\}$ 	& $\{0\}$ \\	
		\end{tabular}
		\hfil
		\begin{tabular}{c|cccc}
			$A\wedge B$ & $\{1\}$ & $\{1, 0\}$ & $\{ \ \}$ & $\{0\}$\\
			\hline
			$\{1\}$ 	& $\{1\}$ 		& $\{1, 0\}$ 	& $\{ \ \}$ & $\{0\}$\\
			$\{1, 0\}$ 	& $\{1, 0\}$ 		& $\{1, 0\}$ 	& $\{0\}$ & $\{0\}$\\
			$\{ \ \}$ & $\{ \ \}$	& $\{0\}$ 	& $\{ \ \}$ & $\{0\}$\\
			$\{0\}$ 	& $\{0\}$ 		& $\{0\}$ 	& $\{0\}$ & $\{0\}$\\	
		\end{tabular}
	\end{center}

\begin{center}
		\begin{tabular}{c|cccc}
			$A\vee B$ & $\{1\}$ & $\{1, 0\}$ & $\{ \ \}$ & $\{0\}$\\
			\hline
			$\{1\}$ 	& $\{1\}$ 	& $\{1\}$ 	& $\{1\}$ 		& $\{1\}$\\
			$\{1, 0\}$ 	& $\{1\}$ 	& $\{1, 0\}$ 	& $\{1\}$ 		& $\{1, 0\}$\\
			$\{ \ \}$ & $\{1\}$ 	& $\{1\}$ 	& $\{ \ \}$ 	& $\{ \ \}$\\
			$\{0\}$ 	& $\{1\}$ 	& $\{1, 0\}$ 	& $\{ \ \}$ 	& $\{0\}$\\	
		\end{tabular}
		\hfil
		\begin{tabular}{c|cccc}
			$A\rightarrow B$ & $\{1\}$ & $\{1, 0\}$ & $\{ \ \}$ & $\{0\}$\\
			\hline
			$\{1\}$ 	& $\{1\}$ 	& $\{1, 0\}$ 	& $\{ \ \}$ 	& $\{0\}$\\
			$\{1, 0\}$ 	& $\{1\}$ 	& $\{1, 0\}$ 	& $\{1\}$ 		& $\{1, 0\}$\\
			$\{ \ \}$ & $\{1\}$ 	& $\{1\}$ 	& $\{ \ \}$ 	& $\{ \ \}$\\
			$\{0\}$ 	& $\{1\}$ 	& $\{1\}$ 	& $\{1\}$ 		& $\{1\}$\\	
		\end{tabular}
	\end{center}

\noindent
Note that $\{1\}$, $\{0\}$, $\{1, 0\}$ and $\{ \ \}$ are not truth values, but assignments or evaluations; more precisely, collections of truth values.\footnote{One might say that they are \emph{generalized truth values}. (See for example \cite{ShramkoWansing2011}.) Let that be. It is still the case that they are not truth values like 1 and 0.} So, under this presentation, it would be rather wrong to call \textbf{FDE} `a four-valued logic'.

Now it can be seen more clearly how three well-known logics, (strong) Kleene logic, \textbf{K$_{3}$}, González-Asenjo/Priest's \textbf{LP} and classical logic can be obtained as extensions of \textbf{FDE}, as I mentioned in the Introduction. \textbf{K$_{3}$} is obtained by ignoring the evaluation $\{1, 0\}$; \textbf{LP} is obtained by ignoring the evaluation $\{ \ \}$; by ignoring those two evaluations at once, one obtains classical logic. Let us move now to the less familiar cases.

\section{Dunn semantics for contra-classical logics}
The evaluation conditions for the binary connectives above have the following general shape

\medskip

\noindent
$1\in\sigma(A\copyright B)$ iff $1\in\sigma(A)$ \emph{copyright} $1\in\sigma(B)$

\medskip

\noindent
$0\in\sigma(A\copyright B)$ iff $v_{i}\in\sigma(A)$ \emph{connective} $0\in\sigma(B)$

\medskip

\noindent
where $v_{i}\in\{1, 0\}$, `\emph{copyright}' stands for a metalinguistic counterpart of $\copyright$ and `\emph{connective}' stands for a metalinguistic counterpart of some other connective, in general distinct from $\copyright$.\footnote{Note that I assume classical reasoning at the meta-theoretical level ---not because I think it is unavoidable but to make things simpler--- and take ``$A$ is false or $B$ is true'' as equivalent to ``If $A$ is true then $B$ is true''. Thus, the truth condition for $A\rightarrow B$ also fits the general shape of truth conditions.} Even more generally, an evaluation condition, regardless of whether it is a truth or falsity condition, has the following general shape:

\begin{center}
$v_{i}\in\sigma(A\copyright B)$ iff $v_{j}\in\sigma(A)$ \emph{relation} $v_{k}\in\sigma(B)$
\end{center}

\noindent
i.e. a value ---truth or falsity-- is assigned to a formula iff there is a certain relation between the assignments of the components.

With this in mind, the changes in an evaluation condition might be of one among the following kinds (the list is not meant to be exhaustive):
\begin{itemize}
    \item C1. The \emph{value} assigned to at least one of the components is changed ---ex. gr. from $1\in\sigma(A)$ to $0\in\sigma(A)$---;
    \item C2. At least one \emph{kind of assignment} is changed ---ex. gr. from $1\in\sigma(A)$ to $1\notin\sigma(A)$---;
    \item C3. The \emph{relation between the assignments} is changed ---ex. gr. from ``$1\in\sigma(A)$ and $0\in\sigma(B)$'' to ``if $1\in\sigma(A)$ then $0\in\sigma(B)$''---;
    \item C4. \emph{Extra conditions are added} ---ex. gr. from ``$1\notin\sigma(A)$ or $0\in\sigma(B)$'' to ``$1\notin\sigma(A)$ or $0\in\sigma(B)$, and $0\notin\sigma(B)$ or $0\in\sigma(A)$''---;
    \item C5. A \emph{mixture} of the above ---ex. gr. from ``$1\in\sigma(A)$ and $0\in\sigma(B)$'' to ``$1\notin\sigma(A)$ or $0\in\sigma(B)$''---.
\end{itemize}

There are special cases of C5, namely mixing C1 and C2, that will be called `tweakings', as they are not so radical changes. All the other changes will be called `modifications'.\footnote{This might go against some uses in the literature, as `tweaking' is used for any kind of change. I try to put some order in the terminology associated to the changes made to the evaluation conditions.} A \emph{Dunn atom} is an expression of the form $v_{i}\in\sigma(A)$ or $v_{j}\notin\sigma(A)$, with $v_{i}, v_{j}\in\{1, 0\}$. Let $v_{i}\in\sigma(A)$ (resp. $v_{j}\notin\sigma(A)$) be a Dunn atom. I will say that $v_{j}\notin\sigma(A)$ (resp. $v_{i}\in\sigma(A)$), with $v_{i}, v_{j}\in\{1, 0\}$ and $v_{i}\neq v_{j}$, is its \emph{Boolean counterpart}. (And I will assume that the relation of being a Boolean counterpart is symmetric.) For instance, the following cases ---horizontal-wise--- are Boolean counterparts of each other:

\begin{center}
\begin{multicols}{2}
$1\in \sigma(\sim A)$

$0\notin \sigma(\sim A)$
\end{multicols}
\end{center}

\begin{center}
\begin{multicols}{2}
$0\in \sigma(A \wedge B)$

$1\notin \sigma(A\wedge B)$
\end{multicols}
\end{center}

\begin{center}
\begin{multicols}{2}
$0\notin \sigma(A \vee B)$

$1\in \sigma(A\vee B)$
\end{multicols}
\end{center}

A \emph{tweaking} is, then, a modification in the evaluation conditions of a connective by changing at least one of its Dunn atoms by their Boolean counterparts in the right-hand side of the `iff'. As an illustration of tweakings, consider negation as evaluated in \textbf{FDE} in the upper left corner and three connectives obtained by changing at least part of its evaluation conditions:

\begin{multicols}{2}
\begin{center}
$1\in \sigma(\sim \! A)$ iff $0\in \sigma(A)$

$0\in \sigma(\sim \! A)$ iff $1\in \sigma(A)$
\end{center}

\begin{center}
$1\in \sigma(\sim_{1} \! A)$ iff $0\in \sigma(A)$

$0\in \sigma(\sim_{1} \! A)$ iff $0\notin \sigma(A)$
\end{center}

\begin{center}
$1\in \sigma(\sim_{2} \! A)$ iff $1\notin \sigma(A)$

$0\in \sigma(\sim_{2} \! A)$ iff $1\in \sigma(A)$
\end{center}

\begin{center}
$1\in \sigma(\sim_{3} \! A)$ iff $1\notin \sigma(A)$

$0\in \sigma(\sim_{3} \! A)$ iff $0\notin \sigma(A)$
\end{center}
\end{multicols}

In fact, $\sim_{3}$ is an old acquaintance: it is \emph{Boolean} negation. Having it available will make presentation easier, and hence I will use a special sign for it: $\neg A$. Another connective that can be added to the language to make presentations easier is the \emph{material} conditional, $A\supset B$, evaluated as follows:

\noindent
$1\in\sigma(A\supset B)$ iff $1\notin\sigma(A)$ or $1\in\sigma(B)$

\noindent
$0\in\sigma(A\supset B)$ iff $1\in\sigma(A)$ and $0\in\sigma(B)$

\noindent
Note that the material conditional can be regarded in its turn as a tweaking on the evaluation conditions for \textbf{FDE}'s conditional.

The truth tables for both Boolean negation and material conditional are as follows:

\begin{center}
		\begin{tabular}{c|c}
			$\neg A$ & $A$ \\
			\hline
			$\{0\}$ 	& $\{1\}$ \\
			$\{ \ \}$	& $\{1, 0\}$ \\
			$\{1, 0\}$ & $\{ \ \}$ \\
			$\{1\}$ 	& $\{0\}$ \\	
		\end{tabular}
		\hfil
		\begin{tabular}{c|cccc}
			$A\supset B$ & $\{1\}$ & $\{1, 0\}$ & $\{ \ \}$ & $\{0\}$\\
			\hline
			$\{1\}$ 	& $\{1\}$ 	& $\{1, 0\}$ 	& $\{ \ \}$ 	& $\{0\}$\\
			$\{1, 0\}$ 	& $\{1\}$ 	& $\{1, 0\}$ 	& $\{ \ \}$ 		& $\{0\}$\\
			$\{ \ \}$ & $\{1\}$ 	& $\{1\}$ 	& $\{1\}$ 	& $\{1\}$\\
			$\{0\}$ 	& $\{1\}$ 	& $\{1\}$ 	& $\{1\}$ 		& $\{1\}$\\	
		\end{tabular}
	\end{center}
	
\medskip

\noindent
A \emph{material biconditional}, $A\equiv B$, can be defined as $(A\supset B)\wedge(B\supset A)$.

\medskip

Let me begin now with the examples of contra-classical logics using the machinery from above. 

\paragraph{Examples: Modifying the evaluation conditions for negation.}

Let all the evaluation conditions fixed, except the truth condition for negation,

\noindent
$1\in\sigma(\sim \! A)$ iff $0\in\sigma(A)$

\noindent
and replace this condition by the following condition

\noindent
$1\in\sigma(\sim \! A)$ iff $0\notin\sigma(A)$

\noindent
which validates $A\vee\sim\sim \! A$ but not $A\vee\sim \! A$. Paul Ruet's \cite{Ruet1996} introduced this connective with the notation $\circlearrowright$, but I will write it `$\sim_{R}$'. Its truth table would be as follows:

\begin{center}
        \begin{tabular}{c|c}
			$\sim_{R} \! A$ & $A$ \\
			\hline
			$\{1, 0\}$ 	& $\{1\}$ \\
			$\{0\}$	& $\{1, 0\}$ \\
			$\{1\}$ & $\{ \ \}$ \\
			$\{ \ \}$ 	& $\{0\}$ \\	
		\end{tabular}
\end{center}

\noindent
Note that the negation so modified is a demi-negation in the sense of Humberstone \cite{Humberstone1995}, since $\sigma(\sim_{R}\sim_{R}~A)$ $=\sigma(\neg A)$, for all $\sigma$, as it can be easily verified.

Now, let all the evaluation conditions fixed, except the falsity condition for negation,

\noindent
$0\in\sigma(\sim \! A)$ iff $1\in\sigma(A)$

\noindent
and replace this condition by the following condition

\noindent
$0\in\sigma(\sim \! A)$ iff $1\notin\sigma(A)$

\noindent
then one obtains Kamide's \cite{Kamide2017} demi-negation of his logic \textbf{CP}, extensively studied in \cite{OmoriWansing2018}, and that I will write as `$\sim_{K}$'. Its truth table would be as follows:

\begin{center}
        \begin{tabular}{c|c}
			$\sim_{K} \! A$ & $A$ \\
			\hline
			$\{ \ \}$ 	& $\{1\}$ \\
			$\{1\}$	& $\{1, 0\}$ \\
			$\{0\}$ & $\{ \ \}$ \\
			$\{1, 0\}$ 	& $\{0\}$ \\	
		\end{tabular}
\end{center}

\noindent
As it is highlighted in \cite{OmoriWansing2018}, \textbf{CP} validates both $\sim_{K} \! (A\wedge\sim_{K} \! \sim_{K} \! A)$ and $\sim_{K} \! \sim_{K} \! (A\wedge\sim_{K} \! \sim_{K} \! A)$ ---which makes it negation-inconsistent---. Note that $\sim_{K}$ is a demi-negation too, since $\sigma(\sim_{K}\sim_{K} A)=\sigma(\neg A)$, for all $\sigma$, yet in general $\sigma(\sim_{R} \! A)\neq\sigma(\sim_{K} \! A)$.

\paragraph{Examples: Modifying the evaluation conditions for conjunction.} Let all the evaluation conditions fixed, except the truth condition for conjunction,

\noindent
$1\in\sigma(A\wedge B)$ iff $1\in\sigma(A)$ and $1\in\sigma(B)$

\noindent
and replace this condition by the following condition

\noindent
$1\in\sigma(A\wedge B)$ iff $1\in\sigma(A)$ or $1\in\sigma(B)$

\noindent
This is a version of \texttt{tonk}: it has the truth condition of disjunction and the falsity condition of conjunction. Defined like that, \texttt{tonk} turned out to be inexpressible in classical logic, but it is in an \textbf{FDE}-like setting. This is its truth table:

\begin{center}
		\begin{tabular}{c|cccc}
			$A\wedge_{t} B$ & $\{1\}$ & $\{1, 0\}$ & $\{ \ \}$ & $\{0\}$\\
			\hline
			$\{1\}$ 	& $\{1\}$ 	& $\{1, 0\}$ 	& $\{1\}$ 		& $\{1, 0\}$\\
			$\{1, 0\}$ 	& $\{1, 0\}$ 	& $\{1, 0\}$ 	& $\{1, 0\}$ 		& $\{1, 0\}$\\
			$\{ \ \}$ & $\{1\}$ 	& $\{1, 0\}$ 	& $\{ \ \}$ 	& $\{0\}$\\
			$\{0\}$ 	& $\{1, 0\}$ 	& $\{1, 0\}$ 	& $\{0\}$ 	& $\{0\}$\\	
		\end{tabular}
\end{center}

\noindent
The resulting logic would be contra-classical, since $\sim \! (A\wedge_{t}\sim \! A)\equiv(A\wedge_{t}\sim \! A)$.\footnote{This also shows that certain classically equivalent definitions of connectives are not so in non-classical contexts. For example, the \texttt{tonk} $\wedge_{t}$ cannot be defined as a connective satisfying both the introduction rules for disjunction and the elimination rules for conjunction, since $A\wedge_{t}B\not\models B$ when $\sigma(A)=\{1\}$ and $\sigma(B)=\{ \ \}$.}

Now let all the evaluation conditions fixed, except the falsity condition for conjunction,

\noindent
$0\in\sigma(A\wedge B)$ iff $0\in\sigma(A)$ or $0\in\sigma(B)$

\noindent
and replace this condition by the following condition

\noindent
$0\in\sigma(A\wedge B)$ iff $0\in\sigma(A)$ and $0\in\sigma(B)$

\noindent
Then one obtains Arieli and Avron's \cite{ArieliAvron1996} \emph{informational meet}, $\otimes$, of their logic \textbf{BL}$_{\supset}$. Following our previous convention, I will write `$\wedge_{AA}$' instead of `$\otimes$'. Its truth table is as follows:

\begin{center}
		\begin{tabular}{c|cccc}
			$A\wedge_{AA} B$ & $\{1\}$ & $\{1, 0\}$ & $\{ \ \}$ & $\{0\}$\\
			\hline
			$\{1\}$ 	& $\{1\}$ 	& $\{1\}$ 	& $\{ \ \}$ 		& $\{ \ \}$\\
			$\{1, 0\}$ 	& $\{1\}$ 	& $\{1, 0\}$ 	& $\{ \ \}$ 		& $\{0\}$\\
			$\{ \ \}$ & $\{ \ \}$ 	& $\{ \ \}$ 	& $\{ \ \}$ 	& $\{ \ \}$\\
			$\{0\}$ 	& $\{ \ \}$ 	& $\{0\}$ 	& $\{ \ \}$ 	& $\{0\}$\\	
		\end{tabular}
\end{center}

\noindent
\textbf{BL}$_{\supset}$ is a contra-classical logic because $\sim \! (A\wedge_{AA} B) \equiv (\sim \! A \wedge_{AA} \sim \! B)$ holds in it.

\paragraph{Examples: Modifying the evaluation conditions for disjunction.} Let all the evaluation conditions fixed, except the truth condition for disjunction,

\noindent
$1\in\sigma(A\vee B)$ iff $1\in\sigma(A)$ or $1\in\sigma(B)$

\noindent
and replace this condition by the following condition

\noindent
$1\in\sigma(A\vee B)$ iff $1\in\sigma(A)$ and $1\in\sigma(B)$

\noindent
This is dual to the \texttt{tonk} presented above; its truth table is as follows:

\begin{center}
		\begin{tabular}{c|cccc}
			$A\vee_{t} B$ & $\{1\}$ & $\{1, 0\}$ & $\{ \ \}$ & $\{0\}$\\
			\hline
			$\{1\}$ 	& $\{1\}$ 	& $\{1\}$ 	& $\{ \ \}$ 		& $\{ \ \}$\\
			$\{1, 0\}$ 	& $\{1\}$ 	& $\{1, 0\}$ 	& $\{ \ \}$ 		& $\{0\}$\\
			$\{ \ \}$ & $\{ \ \}$ 	& $\{ \ \}$ 	& $\{ \ \}$ 	& $\{ \ \}$\\
			$\{0\}$ 	& $\{ \ \}$ 	& $\{0\}$ 	& $\{ \ \}$ 	& $\{0\}$\\	
		\end{tabular}
\end{center}

\noindent
It is the only modified connective for which I have not found a previous appearance in the logic literature, so I will not say anything else about it. And maybe this is so with good reason, as the table is the same as that for informational meet.\footnote{The reader might wonder about other modifications in the truth condition, not exactly the one given above. If they have a suggestion already studied in the literature, I would greatly appreciate it.}

Now let all the evaluation conditions fixed, except the falsity condition for disjunction,

\noindent
$0\in\sigma(A\vee B)$ iff $0\in\sigma(A)$ and $0\in\sigma(B)$

\noindent
and replace this condition by the following condition

\noindent
$0\in\sigma(A\vee B)$ iff $0\in\sigma(A)$ or $0\in\sigma(B)$

\noindent
Then one obtains Arieli and Avron's \cite{ArieliAvron1996} \emph{informational join}, $\oplus$, of their logic \textbf{BL}$_{\supset}$. Again, following our previous convention, I will write `$\vee_{AA}$' instead of `$\oplus$'. The table is as follows:

\begin{center}
     \begin{tabular}{c|c c c c}
        $A\vee_{AA} B$ & \{1\}& \{1,0\}& $\{ \ \}$& \{0\} \\ 
        \hline
        \{1\} &  \{1\} & \{1,0\} & \{1\}& \{1,0\}\\
        \{1,0\}& \{1,0\}&  \{1,0\} &\{1,0\}&\{1,0\}\\
        $\{ \ \}$&   \{1\}&  \{1,0\}   &$\{ \ \}$&\{0\}\\
        \{0\}&  \{1,0\}&   \{1,0\} &\{0\}&\{0\}
    \end{tabular}
\end{center}

\noindent
Contra-classicality enters \textbf{BL}$_{\supset}$ not only through $\wedge_{AA}$, but also with this connective since $\sim \! (A\vee_{AA} B) \equiv (\sim \! A \vee_{AA} \sim \! B)$ holds in it. With this one can also show that \textbf{BL}$_{\supset}$ is negation-inconsistent, since it validates both $((A\supset A)\vee_{AA} \sim \! (A\supset A))$ and $\sim \! ((A\supset A)\vee_{AA} \sim \! (A\supset A))$.

\paragraph{Examples: Modifying the evaluation conditions for the conditional.} Let all the evaluation conditions fixed, except the truth condition for the conditional,

\noindent
$1\in\sigma(A\rightarrow B)$ iff if $0\in\sigma(A)$ or $1\in\sigma(B)$

\noindent
and replace it by the following condition

\noindent
$1\in\sigma(A\rightarrow B)$ iff $1\in\sigma(A)$ and $1\in\sigma(B)$

\noindent
then one obtains a logic with the four valued generalization of a connective that has been studied several times in the recent history of logic although in Kleene-like three valued logics: it is Reichenbach's \cite{Reichenbach1935} and \cite{Reichenbach1944} \emph{quasi-implication}, de Finetti's \cite{deFinetti1936} conditional  (or, if not identical, at least very similar to it), and Blamey's \cite{Blamey1986} \emph{transplication}. This is its truth table:

\begin{center}
\begin{tabular}{c|cccc}
			$A\rightarrow_{DF} B$ & $\{1\}$ & $\{1, 0\}$ & $\{ \ \}$ & $\{0\}$\\
			\hline
			$\{1\}$ 	& $\{1\}$ 	& $\{1, 0\}$ 	& $\{ \ \}$ 	& $\{0\}$\\
			$\{1, 0\}$ 	& $\{1\}$ 	& $\{1, 0\}$ 	& $\{ \ \}$ 		& $\{0\}$\\
			$\{ \ \}$ & $\{ \ \}$ 	& $\{ \ \}$ 	& $\{ \ \}$ 	& $\{ \ \}$\\
			$\{0\}$ 	& $\{ \ \}$ 	& $\{ \ \}$ 	& $\{ \ \}$ 		& $\{ \ \}$\\	
		\end{tabular}
\end{center}

\noindent
It validates, among other things, $(A\rightarrow_{DF} B)\models A$.\footnote{See \cite{EgreRossiSprenger2020} for a recent comprehensive study of such connective in three-valued settings.}

Now consider an expansion of \textbf{FDE} with the material conditional and let all the evaluation conditions fixed, except the falsity condition for the material conditional,

\noindent
$0\in\sigma(A\supset B)$ iff $1\in\sigma(A)$ and $0\in\sigma(B)$

\noindent
and replace it by the following condition

\noindent
$0\in\sigma(A\supset B)$ iff $1\notin\sigma(A)$ or $0\in\sigma(B)$

\noindent
then one obtains \textbf{MC}, `material connexive logic', introduced in \cite{Wansing2020}. The truth table for such conditional is the following one:

\begin{center}
\begin{tabular}{c|cccc}
			$A\rightarrow_{W} B$ & $\{1\}$ & $\{1, 0\}$ & $\{ \ \}$ & $\{0\}$\\
			\hline
			$\{1\}$ 	& $\{1\}$ 	& $\{1, 0\}$ 	& $\{ \ \}$ 	& $\{0\}$\\
			$\{1, 0\}$ 	& $\{1\}$ 	& $\{1, 0\}$ 	& $\{ \ \}$ 		& $\{0\}$\\
			$\{ \ \}$ & $\{1, 0\}$ 	& $\{1, 0\}$ 	& $\{1, 0\}$ 	& $\{1, 0\}$\\
			$\{0\}$ 	& $\{1, 0\}$ 	& $\{1, 0\}$ 	& $\{1, 0\}$ 		& $\{1, 0\}$\\	
		\end{tabular}
\end{center}

\noindent
Connexive logics including such a conditional typically validate both Boethius' Thesis and its converse, i.e. \emph{Wansing's Thesis} $\sim \! (A\rightarrow_{W} B)\leftrightarrow_{W} (A\rightarrow_{W} \sim \! B)$.\footnote{Richard Sylvan \cite{Sylvan1989BG4} dubbed `hyper-connexivism' the thesis that, in addition to Aristotle's and Boethius' Theses, the converses of the latter also hold.} They are also negation-inconsistent; as witnesses, take $(A\wedge \sim \! A)\rightarrow_{W} A$ and $\sim((A\wedge \sim \! A)\rightarrow_{W} A)$.\footnote{A connexive variant of the more general version of Belnap-Dunn logic, including the negation $\neg$, was studied in \cite{Omori2016} under the name `\textbf{dBD}'.}

\paragraph{Example: Modifying several evaluation conditions at the same time.} If one modifies at least two falsity conditions, for example, that for the conditional as in the connexive logic above, that for disjunction as in \textbf{BL}$_{\supset}$, and the falsity condition for conjunction so that one gets

\noindent
$0\in\sigma(A\wedge B)$ iff either $1\in\sigma(A)$ and $0\in\sigma(B)$, or $0\in\sigma(A)$ and $1\in\sigma(B)$

\noindent
then one obtains Francez's \cite{Francez2019a} \textbf{PCON}, which inherits the contra-classical features of the logics on which it is based.\footnote{`\textbf{PCON}' stands for `poly-connexive logic'. It is \emph{multi-contra-classical} in the sense that it validates contra-classical theses for more than two connectives.} (Although, as the peculiar falsity condition for conjunction witnesses, this logic was motivated by considerations independent from  of those by Avron and Arieli.) 

I have not found in the literature a contra-classical logic that can be described as a variant of \textbf{FDE} in which the truth conditions for two or more connectives are modified. I would greatly appreciate suggestions on this regard.

\section{The source of contra-classicality}
At this point, at least two questions can be asked:

\noindent
(Q1) What is the connection between contra-classicality and the evaluation conditions such that certain modifications in the latter produce the former?

\noindent
(Q2) Are the modified connectives still the intended connectives?

\noindent
Perhaps the second is the more pressing one. It is far from clear that, say, Kamide's demi-negation is still a negation, or whether the de Finetti connective is still a conditional, to mention just two examples of the difficulty. Nonetheless, I will leave it for another occasion. Tackling at least the first one is already an important contribution.\footnote{Nonetheless, see again \cite{OmoriWansing2018} for the question whether Kamide's connective is a negation, and \cite{EgreRossiSprenger2020} and \cite{EstradaRamirez202X} for discussion about the conditionality of the de Finetti conditional.}

As we have seen above, contra-classicality is not restricted to modifying just one of the evaluation conditions, either the truth or the falsity condition; contra-classical logics can be obtained by modifying either of them. As the history of connexive logic witnesses, most contra-classical logics have been obtained by modifying the truth conditions alone. In the case of connexive logics, it has been the truth condition for the conditional (combined with Boolean negation rather than de Morgan negation); see for example the connexive logics so obtained in \cite{Angell1962}, \cite{McCall1966}, \cite{Pizzi1977}, \cite{Pizzi1991} and \cite{Priest1999}. And a recent wave of connexive logics, after \cite{Wansing2005} ---such as \cite{Omori2016} and \cite{WansingUnterhuber2019}--- modify the falsity condition (using the de Morgan negation).\footnote{In all fairness, connexive logics have been obtained by other means than model-theoretically. See for example \cite{RahmanRuckert2001} or \cite{McCall2014}, for proof theoretic-based connexive logics. In fact, as a referee correctly points out, this sort of presentation of logics would allow for further extensions of the Bochum Plan, in considering systematic and controlled modifications in the proof-theoretic machinery.}

Moreover, not any change in the evaluation conditions, not even a large number of them, produces contra-classical logics. For example, if one tweaks the falsity conditions of all of $\sim$, $\wedge$, $\vee$ and $\rightarrow$ as follows

\noindent
$0\in(\sim \! A)$ iff $0\notin\sigma(A)$

\noindent
$0\in\sigma(A\wedge B)$ iff either $1\notin\sigma(A)$ or $1\notin\sigma(B)$

\noindent
$0\in\sigma(A\vee B)$ iff $1\notin\sigma(A)$ and $1\notin\sigma(B)$

\noindent
$0\in\sigma(A\rightarrow B)$ iff $0\notin\sigma(A)$ and $1\notin\sigma(B)$

\noindent
one obtains a four-valued generalization of Sette's \textbf{P}$^1$, which is not contra-classical.\footnote{It would be worth comparing this generalization with those studied in \cite{Omori2017}. This is left for another work.}

What kind of modifications does the job then? Although it remains to be properly proved, the examples strongly suggest that only those modifications that make one of the evaluation conditions classically equivalent to the corresponding classical evaluation condition for some other connective deliver contra-classicality. In the examples on Section 3,

\begin{itemize}
\item the modified truth condition for negation is (classically) that of identity;
\item the modified falsity condition for negation is (classically) that of identity;
\item the modified truth condition for conjunction is (classically) that of disjunction;
\item the modified falsity condition for conjunction is (classically) that of disjunction;
\item the modified truth condition for disjunction is (classically) that of conjunction;
\item the modified falsity condition for disjunction is (classically) that of conjunction;
\item the modified truth condition for the conditional is (classically) that of conjunction;
\item the modified falsity condition for the conditional is (classically) that of conjunction;
\end{itemize}

It is in this case of modified evaluation conditions where contra-classicality enters the scene: the modified evaluation condition endows the connective with properties of some other connective, and hence validates things that it does not validate in classical logic.

\section{Conclusions}
In this paper, by using Dunn semantics I gave systematic changes in the evaluation conditions for negation, conjunction, disjunction and the conditional, and relate most of them with already existing contra-classical logics. This means that those contra-classical logics can be regarded as variants of \textbf{FDE}, obtained by modifying the evaluations conditions for certain connectives. Then I argued that such systematization provides a clear understanding of the source of contra-classicality, namely, connectives that have either the classical truth or the classical falsity condition of another connective.

A pressing question remains open at this point: are the modified connectives still the intended connectives? Why? This is, as I have said, left for future work. But there are at least two more paths to follow after this investigation. First, other standard connectives, like the 0-ary connectives or constants, some modal connectives or the usual quantifiers, can be given Dunn semantics. Finding examples of contra-classical logics involving those other connectives might be instructive as well to test the explanatory power that the Dunn semantics seems to possess. Second, the entailment relation was assumed to be Tarskian. But it has recently been argued ---see \cite{EgreRossiSprenger2020}, \cite{EstradaRamirez202X}--- that certain connectives, like transplication, are closer to its intended connective when the entailment relation is not Tarskian, in particular, when it is non-transitive. This suggests that the underlying notion of entailment would be allowed to vary as well, not only the truth and falsity conditions, which would make the space of contra-classical logics even richer. This idea also deserves a systematic exploration.

\bibliographystyle{eptcs}
\bibliography{biblio}
\end{document}